\documentclass[prl,showpacs,amsmath,aps,twocolumn,superscriptaddress]{revtex4-1}
\usepackage{amsfonts}
\usepackage{graphicx,color}
\begin{document}

\title{MeV Dark Matter in light of the Small Scale Crisis}

\author{Changhong~Li}
\email{changhongli@ynu.edu.cn}
\affiliation{Department of Astronomy,  Key Laboratory of Astroparticle Physics of Yunnan Province, School of Physics and Astronomy,  Yunnan University, No.2 Cuihu North Road, Kunming, China 650091}

\begin{abstract}
The small-scale crisis is one of the most outstanding puzzles in modern cosmology and astrophysics. It may imply a suppression of matter perturbation at small scale. In this work, by taking into account of the gravitational effects from the non-equilibrium production of DM, we propose a new mechanism, which can realize such desired suppression and alleviate the crisis within the framework of cold dark matter (DM) and simple inflation. Moreover, in this new mechanism, we establish a novel relation between the particle mass of DM, $m_\chi$, and the critical scale of suppression, $k_\star^{-1}$. As  $k_\star^{-1}$ will be further constrained in future astrophysical observations, $m_\chi$ can be constrained accordingly with this relation. It thus provides a new method in complementary to other existing strategies of determining $m_\chi$. Furthermore, to illustrate our theoretical prediction, we consider a suppression at $k_\star^{-1}=1~\text{kpc}$ that can partially alleviate the small-scale crisis, and obtain $m_\chi=2.2~\text{MeV}$ for realizing such suppression. Then we plot the power spectrum of linear matter perturbation for this case, and illustrate a salient feature of the suppression that can serves a smoking-gun signature of this new mechanism in future observations.

\end{abstract}


\maketitle



\section{Introduction} 
The small-scale crisis, which describes the discrepancies on the sub-galactic scale structure formation amongst the observations, theoretical predictions and N-body simulations, is one of the most outstanding puzzles in modern cosmology and astrophysics \cite{Weinberg:2013aya}. Since these discrepancies, which include the missing-satellite problem, the cusp-vs-core problem and the too-big-to-fail problem \cite{Moore:1999nt, Moore:1999gc, BoylanKolchin:2011de, Nakama:2017ohe}, seem hard to be reconciled with the two basic assumptions: 1) dark matter (DM) is cold and 2) the inflation is simple, they are, collectively, highlighted as a crisis. Perhaps, such crisis will be alleviated with further elucidation of baryonic physics and relevant astrophysical observations (for example, see \cite{Kim:2017iwr,Koposov:2015cua,Drlica-Wagner:2015xua,Simon:2007dq}). Besides it, in this work, we propose a new mechanism, which can realize a suppression of matter perturbation to alleviate this crisis within the framework of simple inflation and cold DM. 

\begin{figure}[htp!]
\centering
\includegraphics[width=0.48\textwidth]{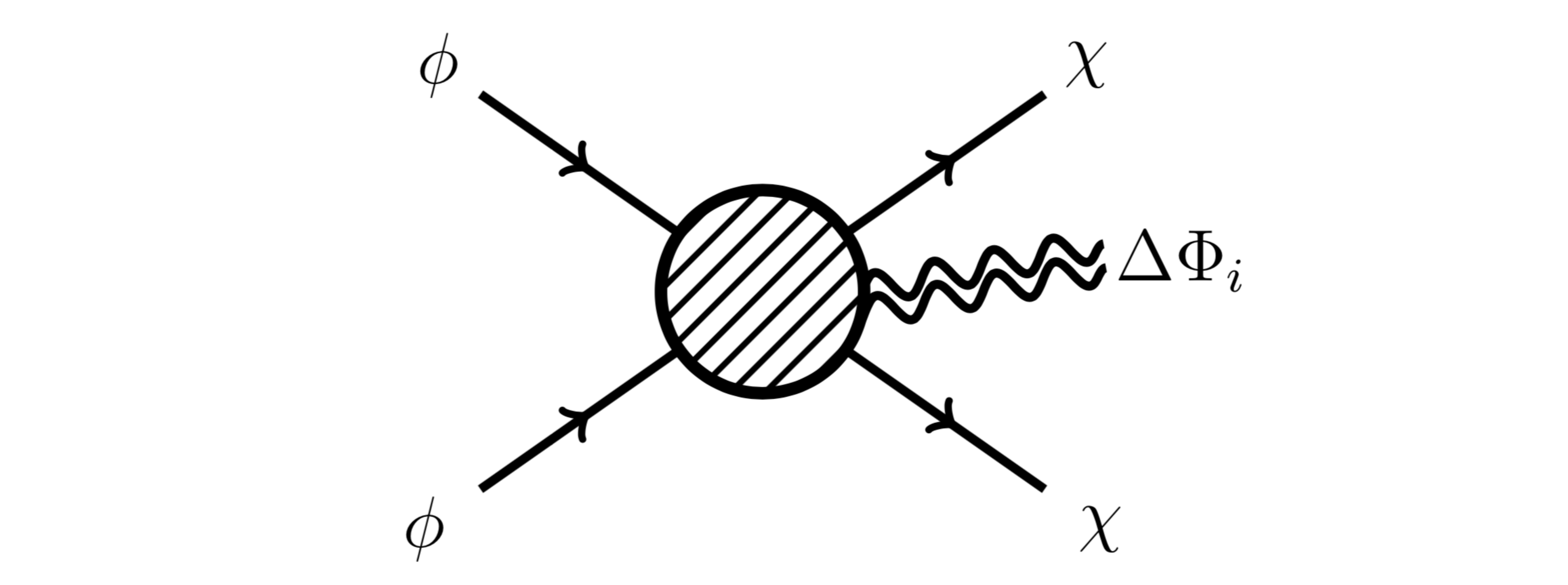}
\caption{A schematic plot for a pair production of DM particles $\chi$ via the pair annihilation of scalar particles, $\phi$. In curved and nonrigid spacetime, a small local fluctuation of spacetime, $\Delta\Phi_i$, can be generated ({\it c.f.} FIG.1.b in \cite{Li:2019cjp}) in each pair production of DM particles. }
\label{fig:feyman.pdf}
\end{figure}

Specifically, we take into account of the resonant gravitational effects from the non-equilibrium production of cold DM, and demonstrate that it can lead to a desired suppression of matter perturbation. As illustrated in FIG.\ref{fig:feyman.pdf}, in the early Universe, each pair production of DM particles can imprint a small local fluctuation, $\Delta\Phi_i$, in the trace of metric. As demonstrated in Refs.\cite{Li:2019cjp} and \cite{Li:2019std}, during DM production, such fluctuations can accumulate to drive a resonance between the DM density perturbation and the scalar modes of metric perturbation, which can amplify both of them. Specifically, only the long wavelength modes of metric and density perturbations get full amplification. For the short wavelength modes, they however get less amplification since they were re-entering horizon before the end of DM production. Such difference in amplifications thus leads to a relative suppression of the linear metric and matter perturbations at small scales \footnote{Note that, since the primordial gravitational wave has not been detected, an amplification of the scalar modes of primordial metric perturbation and linear matter perturbation at large scales only implies a relative suppression at small scales, see \cite{Li:2019cjp} for more details.}, and can, in part, alleviate the small-scale crisis.   

Notably, this new proposal is alternative to other well-known suppression mechanisms, such as the {\it primordial suppression}  and the {\it late-time suppression}. In the former \cite{Kamionkowski:1999vp,Yokoyama:2000tz, Kobayashi:2010pz}, a non-simple inflation, which can break the scale-invariance of primordial curvature perturbation spectrum, is employed to realize the suppression  of the matter perturbation at small scale. And the suppression takes place during the inflation. In the latter \cite{Bode:2000gq,Viel:2011bk, Lin:2000qq,Sigurdson:2003vy,Kusenko:2009up,Rocha:2012jg,Peter:2012jh, Hochberg:2014dra, Foot:2014uba, Foot:2016wvj}, they take non-cold (warm or new exotic interacting) DM candidates to realize the suppression, and the suppression happens only after the horizon re-entering. What notable for our new mechanism is that, the suppression can be realized with simple inflation and cold DM. Moreover, such suppression takes place in a unique epoch, from the end of inflation to the horizon re-entering. To avoid confusion, we call it ``{\it early-time suppression}'' mechanism.     

In this work, after briefly reviewing the non-equilibrium production of non-thermal DM, we elaborate the early-time suppression mechanism, and compute the critical scale of the suppression of matter perturbation, $k_\star^{-1}$. We then obtain a novel relation between $k_\star^{-1}$ and the particle mass of DM, $m_\chi$, which can be used to constrain $m_\chi$ with the future observations on $k_\star^{-1}$. It thus provides a new method to determine $m_\chi$ with the astrophysical observation. Furthermore, to illustrate our theoretical prediction, we consider a suppression at $k_\star^{-1}=1~\text{kpc}$ that can partially alleviate the small-scale crisis, and by using the newly obtained relation, we obtain $m_\chi=2.2~\text{MeV}$ for realizing such suppression. Then we plot the power spectrum of primordial linear matter perturbation for this case to illustrate a salient feature of the early-time suppression mechanism, the smaller the scale is, the stronger the suppression is. As the location and shape of small scale suppression of matter perturbation will be precisely measured in future astrophysical observations  (and for recent developments of observational constraints on the matter perturbation spectrum see \cite{Irsic:2017ixq, Yeche:2017upn, Viel:2013apy, Palanque-Delabrouille:2013gaa, Dawson:2012va}, and references there in), such feature can serve as a smoking-gun signature for testifying our early-time suppression mechanism.

\section{Non-equilibrium Production of DM and its Gravitational effects} 
Following Ref.\cite{Li:2019cjp}, we consider a simple realization of the non-thermal DM scenario \cite{Baer:2014eja,Chung:1998ua,Shi:1998km, Feldstein:2013uha, Li:2014era, Cheung:2014nxi}. In which, DM particles, $\chi$, are produced by the pair annihilations of the light scalar particles, $\phi$, with the minimal coupling, $\mathcal{L}=\lambda \phi^2\chi^2$. Due to its small cross-section, even after cosmic reheating, the production of non-thermal DM particles will last for a long time until it freezing out. And during this process of non-thermal DM production, its abundance is always out of the thermal equilibrium, $Y_\chi\ll Y_\chi^{eq}$, where $Y_\chi\equiv n_\chi T^{-3}$ and $n_\chi$ are, respectively, the abundance and number density of DM particles, and the superscript $^{eq}$ denotes the thermal equilibrium. What notable is that, during such non-equilibrium production of non-thermal DM, the fluctuation of DM density will not be suppressed or washed out by the envelopment of thermal equilibrium. We therefore can study the nontrivial evolution of DM density perturbation during this phase and investigate its gravitational effects, which may be probed in astrophysical observations.   

More specifically, during the non-equilibrium production of non-thermal DM, the small local fluctuations of spacetime, $\Delta\Phi_i$ (as shown in FIG.\ref{fig:feyman.pdf}), will accumulate to drive a resonance between DM density perturbation and scalar modes of metric perturbation \cite{Li:2019cjp}. And such resonance can amplify the super-horizon scalar modes of metric perturbation with the added value, $\Delta\Phi=\sum_i\Delta\Phi_i+\Delta\Phi_b$, where $\Delta\Phi_b$ is contributed from the back-reaction of DM density perturbation. By following Ref.~\cite{Li:2019cjp}, we can obtain the evolution of the super-horizon scalar modes of metric perturbation, $\Phi(y)=\Phi_\varphi+\Delta\Phi$, during DM production and after freezing-out({\it c.f.} Eq.(8) in \cite{Li:2019cjp}), 
\begin{equation}\label{eq:phiyevs}
\Phi(y)=
\left\{  
  \begin{array} {lr}
 {\displaystyle \Phi_\varphi \mathcal{G}\left(-y/\xi\right),\qquad ~~  y_{R_f}\le y\le y_f} 
 \\ 
  {\displaystyle  \Phi_\varphi \mathcal{G}\left(-y_f/\xi\right)[1+\mathcal{A}(1-(y_f/y)^{\frac{5}{2}})], y> y_f}   
  \\
\end{array}     
\right.,
\end{equation}
where $y\equiv m_\chi/T$ with $T$ being the background temperature, $\mathcal{G}(x)\equiv {_2F_1}\left(\frac{3-\sqrt{17}}{4},\frac{3+\sqrt{17}}{4};\frac{7}{2}; x \right)$ is Gauss hypergeometric function, and the coefficient $\mathcal{A}\simeq 0.11$. To obtain Eq.(\ref{eq:phiyevs}), the perturbed FLRW metric in conformal Newtonian gauge, $ g_{\mu\nu}=\{-1-2\Psi(\vec{x},t), a^2(t) \delta_{ij} \left[1+2\Phi(\vec{x},t)\right]\}$, is adopted and $\Phi(y)$ denotes the super-horizon Fourier modes of $\Phi(\vec{x},t)$. $\Phi_\varphi$ is the primordial value of $\Phi(y)$ at the end of reheating. $y_{R_f}$ and $y_{f}$ label, respectively, the beginning and the end of the post-reheating DM production. $\xi$ is a dimensionless parameter introduced by Ref.\cite{Li:2019cjp} to characterize how Universe is reheated, and its parameter region takes $y_{R_f}\ll \xi\le1$. Specifically, for a given value of $m_\chi$, a smaller $\xi$ implies a shorter duration of reheating (see Ref.\cite{Li:2019cjp} for more details). In this work, we can simply take it as a constant free parameter.

\begin{figure}[htp!]
\centering
\includegraphics[width=0.48\textwidth]{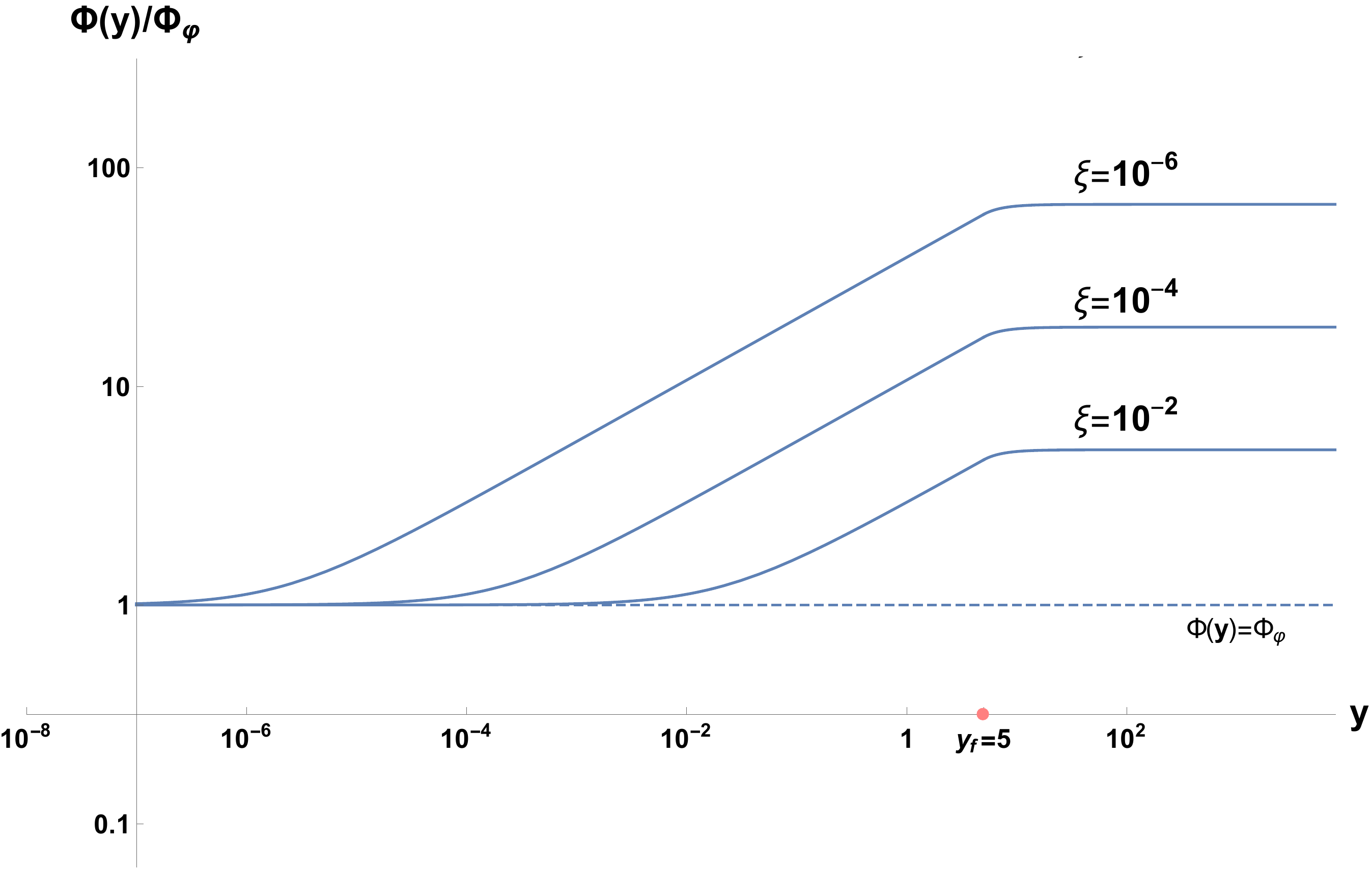}
\caption{The evolution of $\Phi(y)$ during and after DM production.}
\label{fig:Phis.pdf}
\end{figure}

In FIG.\ref{fig:Phis.pdf}, we use Eq.(\ref{eq:phiyevs}) to plot $\Phi(y)$ for the typical cases $\xi=(10^{-2}, 10^{-4}, 10^{-6} )$ with $y_{R_f}=10^{-7}$. This plot shows that,  $\Phi(y)$ is amplified during DM production, the amplification is frozen at the end of DM production ($y=y_f=5$ \footnote{An order-of-magnitude estimate for the ending of freezing-out can give that $1\le y_f\le 10$. For simplicity, we take $y_f=5$ throughout this work.}), and a smaller $\xi$ leads to a larger amplification\footnote{For the physical interpretation of these three features, please see Refs.\cite{Li:2019cjp} and \cite{Li:2019std}. }. In this work, we primarily focus on the difference in the amplifications of the long and the short wavelength modes. For the long wavelength modes, they are out of horizon during the whole process of DM production, and get full amplification. However, for the short wavelength modes, they were out of horizon initially, but after a short while, they are re-entering the horizon before the end of DM production, so that they are getting less amplification. Such difference in amplifications thus results in a relative suppression of $\Phi(y)$ at small scales.   

\section{A novel relation between $m_\chi$ and $k_\star^{-1}$} 

\begin{figure}[htp!]
\centering
\includegraphics[width=0.48\textwidth]{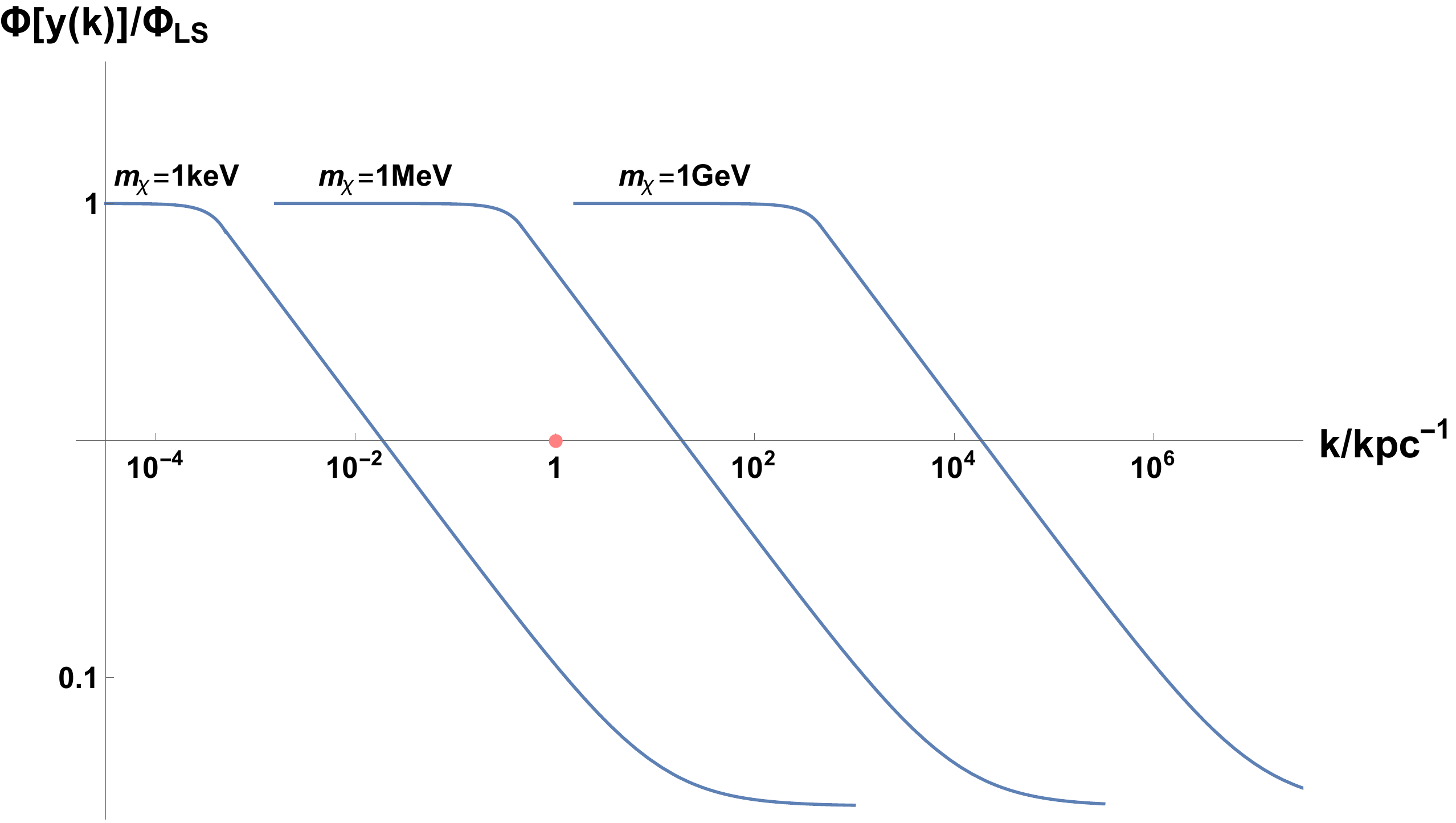}
\caption{$\Phi[y(k)]$ for $m_\chi=\{1\text{keV},1\text{MeV},1\text{GeV}\}$. }
\label{fig:Phik.pdf}
\end{figure}

In order to determine the spatial scale of the suppression of $\Phi(y)$, we can translate $\Phi(y)$ into $\Phi[y(k)]$ by using the horizon-crossing condition,
\begin{equation} \label{eq:horeco}
k\eta(k)=1~,
\end{equation} 
where $\eta(k)$ denotes the conformal time of horizon re-entering for the perturbation mode with wave-vector $k$. During this radiation-dominated era, we have $\eta(k)=c_\ast y(k)m_\chi^{-1}$, where the coefficient $c_\ast=6.88\times 10^{31}$ is constrained by the current observations \footnote{In radiation-dominated era, we have $da=-c_2T^2dT$ and $da=c_1 d\eta$, which lead to $c_1(\eta-\eta_{R_f})=c_2(T^{-1}-T_{R_f}^{-1})$, where $c_1\equiv \left(8\pi G\rho_{e}a_{e}^4/3\right)^{\frac{1}{2}}$ and $c_2= 0.32 a_0 T_0$, and the subscripts $_{R_f}$, $_e$ and $_0$ label the quantities at the end of reheating, the matter-radiation equality, and today respectively.  $\eta_{R_f}$ and $T_{R_f}^{-1}$ are negligible in the parameter region of our interest, so we can get $\eta=c_\ast T^{-1}=c_\ast ym_\chi^{-1}$, where $c_\ast\equiv c_2/c_1=6.88\times 10^{31}$ following \cite{Dodelson:2003ft}.}. By substituting the expression of $\eta(k)$ into Eq.(\ref{eq:horeco}), we obtain
\begin{equation}\label{eq:ymckn}
y(k)=\frac{m_\chi}{c_\ast k}~.
\end{equation} 
Then, by substituting Eq.(\ref{eq:ymckn}) into Eq.(\ref{eq:phiyevs}), we can obtain the expression of $\Phi[y(k)]$. In FIG.\ref{fig:Phik.pdf}, we take $\xi=10^{-4}$ to plot $\Phi[y(k)]$ with $m_\chi=\{1\text{keV},1\text{MeV},1\text{GeV}\}$ \footnote{In this work, $\xi$ is taken a free parameter. In further study, the value of $\xi$ can be computed by employing a specific reheating process. }, where $\Phi[y(k)]$ is normalized by its amplitude at large scale, $\Phi_{LS}\equiv \Phi[y(k=10^{-3}\text{Mpc}^{-1})]$. 

As shown in FIG.\ref{fig:Phik.pdf}, given the value of $\xi$, the amplifications are the same for all large scales. Therefore, the scale-invariance of curvature perturbation spectrum can be preserved by such amplification. However, at small scale, the production of DM can result in a relative suppression of $\Phi[y(k)]$. And we can notice that the larger the $m_\chi$ is, the smaller the critical scale of suppression is. In particular, for $m_\chi= 1~\text{MeV}$, the suppression of $\Phi[y(k)]$ takes place below the kpc scale \footnote{For $m_\chi \simeq 1\text{GeV}$, $k_\star\simeq 0.5\text{pc}^{-1}$, and for $m_\chi \simeq 1\text{keV}$, $k_\star\simeq 0.5\text{Mpc}^{-1}$. The latter is similar to the prediction from the well-known warm DM hypothesis. However, they have different physics origins. For warm DM, the suppression is due to its long free-streaming length. And for this new proposal, the suppression is caused by the pair production of DM particles.}. 

Analytically, to determine the critical scale of suppression, $k_\star^{-1}$, we can simply substitute $y|_{k=k_\star}=y_f$ into Eq.(\ref{eq:ymckn}), and obtain
\begin{equation} \label{eq:mchiandki}
k_\star^{-1}=cy_f m_\chi^{-1},
\end{equation}
which indicates that, while the magnitude of the suppression is determined by the parameter $\xi$, the location of the suppression is solely determined by $m_\chi$. As $k_\star^{-1}$ will be further constrained in the future astrophysical observations, the early-time suppression can be testified and $m_\chi$ can be determined accordingly. 
    
\section{Linear Matter Perturbation Spectrum}
\begin{figure}[htp!]
\centering
\includegraphics[width=0.48\textwidth]{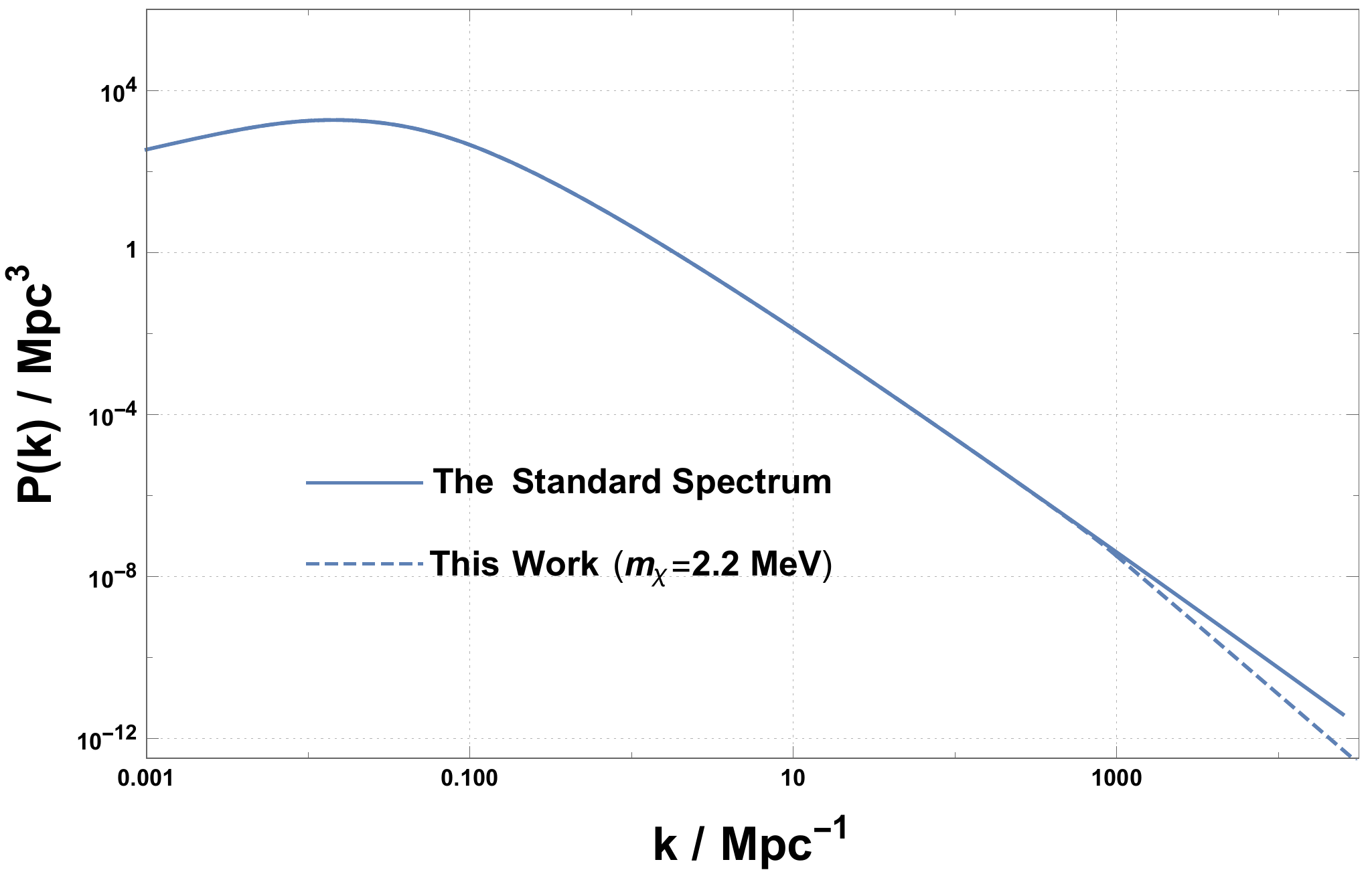}
\caption{The power spectrum of primordial linear matter perturbation are plotted for 1) the standard approach without considering the gravitational effects from DM production (solid line), and 2) the early-time suppression mechanism proposed in this work (dashed line, $\xi=10^{-4}$ and $y_f=5$).}
\label{fig:llppower.pdf}
\end{figure}

In this section, we are plotting the power spectrum of primordial linear matter perturbation for the early-time suppression. For illustration, we take a suppression at $k_\star^{-1}=1~\text{kpc}$, which can partially alleviate the small-scale crisis. By substituting $k_\star^{-1}=1~\text{kpc}$ and $y_f=5$ into Eq.(\ref{eq:mchiandki}), we obtain  
\begin{equation}\label{eq:mccrin}
m_\chi=y_f c_\ast k_\star=2.2~\text{MeV},
\end{equation}
with a free-streaming length, $l=112(m_\chi/\text{eV})^{-1}~\text{Mpc}=0.05~\text{kpc}$ \cite{Peacock:2003hh}. Since its free-streaming length is much smaller than the critical scale of suppression, $k_\star^{-1}=1~\text{kpc}$, we can confirm that, in light of the small-scale crisis, such DM candidate is cold. And such suppression is resulting from the non-equilibrium production of cold non-thermal DM rather than the smoothing mechanism of warm or exotic DM candidates. In FIG.\ref{fig:llppower.pdf}, we adopt the BBKS transfer function for cold DM ({\it c.f.} Eq.(7.70) in \cite{Dodelson:2003ft}), $\mathcal{T}(x\equiv k/\Omega_m h^2 \text{Mpc}^{-1})=\frac{\ln(1+2.34x)}{2.34x}\times\left[1+3.89x+(16.2x)^2+(5.74x)^3+(6.71x)^4\right]^{-0.25}$, to plot the power spectrum of primordial linear matter perturbation, $\mathcal{P}(k)=Ck^{n_s}\mathcal{T}^2(x)(\mathcal{P}_{\Phi[y(k)]}/\mathcal{P}_{\Phi_{LS}})$, where $C=2.69\times 10^5 \text{Mpc}^{3+n_s}$\footnote{$C$ is obtained by normalizing the spectrum on large scale $P(k=10^{-3}\text{Mpc}^{-1})=343\text{Mpc}^{3}$ following Ref.\cite{Dodelson:2003ft}.}, $n_s=0.96$, and $\Omega_m h^2=0.15$ \cite{Komatsu:2010fb, Ade:2015xua}.   

\begin{figure}[htp!]
\centering
\includegraphics[width=0.48\textwidth]{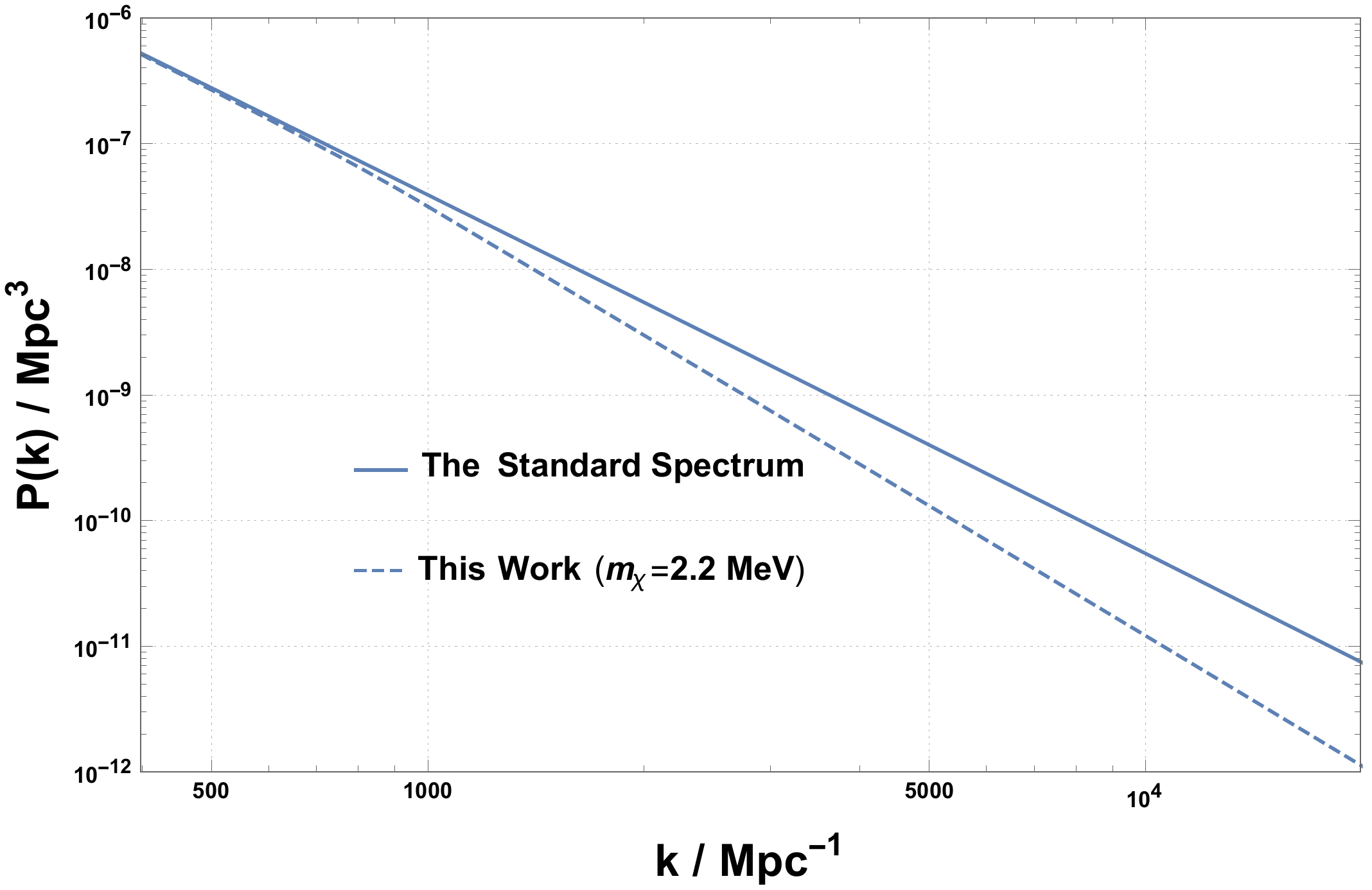}
\caption{High resolution plot for small scale region of FIG.\ref{fig:llppower.pdf}. }
\label{fig:highresolution.pdf}
\end{figure}

As shown in FIG.\ref{fig:llppower.pdf}, by taking into account of the gravitational effects from DM production, we have realized a suppression of linear matter perturbation at small scale within the framework of cold DM and simple inflation. In FIG.\ref{fig:highresolution.pdf}, we re-plot FIG.\ref{fig:llppower.pdf} with a higher resolution. This hi-resolution plot illustrates a salient feature of the early-time suppression, the smaller the scale is, the stronger the suppression is. This feature is, quantitatively, different from the existing proposals such as warm DM candidates. It therefore can serve a smoking-gun signature for testifying our mechanism with  future astrophysical observations, and can be potentially used for extrapolating a small portion of observational data to determine $k_\star^{-1}$ and $m_\chi$ as well. 

\section{Summary} 
In this work, by taking into account of the gravitational effects from the non-equilibrium production of the non-thermal DM, we proposed a new mechanism to alleviate the small-scale crisis through realizing a suppression of matter perturbation at small scales. The novelty of this new mechanism is that the desired suppression can be realized within the framework of simple inflation, cold DM and General Relativity, which is alternative to existing suppression mechanisms that employed non-simple inflation, warm or exotic DM, or modified gravity \cite{Kroupa:2018kgv}. 

Moreover, by computing the location and shape of the suppression in this early-time suppression mechanism, we established a new relation between the particle mass of DM, $m_\chi$, and the critical scale of the suppression, $k_\star^{-1}$, {\it i.e.} Eq.(\ref{eq:mchiandki}). This relation implies that DM candidates with different particle mass can result in a suppression at different small scales. On the other side, as $k_\star^{-1}$ will be further constrained in the development of the astrophysical observations, this relation can be used to determine $m_\chi$ accordingly. It is worthy to emphasize that, as a new method of determining $m_\chi$, this relation established from the perturbative gravitational effects is complimentary to other existing strategies \cite{Patrignani:2016xqp}, which primarily focus on DM abundance.  And this method is also complimentary to the proposal in Refs.\cite{Li:2019cjp} and \cite{Li:2019std}. In which, they mainly focus on the tensor-to-scalar ratio of metric perturbation at large scales, and explore a new way to constrain $m_\chi$ with the primordial gravitational wave searches.

Furthermore, to illustrate our theoretical prediction, we considered a suppression at $k_\star^{-1}=1~\text{kpc}$, which can partially alleviate the small-scale crisis, and plotted its power spectrum of primordial linear matter perturbation. For this case, we obtain $m_\chi=2.2~\text{MeV}$. And our plot also illustrates a salient feature for the early-time suppression, the smaller the scale is, the stronger the suppression is (see FIG.\ref{fig:highresolution.pdf}). As this feature is, quantitatively, different from the predictions of other existing mechanisms, it can serve as a unique signature for testifying the early-time suppression mechanism. Moreover, if only a small portion of suppression is measured in the near future, this feature can be also used to  extrapolate the observational data to determine $k_\star$, and then $m_\chi$.  

At the end, we highlight an issue worthy of further study. In FIG.\ref{fig:Phis.pdf}, we illustrated that the magnitude of suppression is determined by the value of $\xi$. However, in this work, $\xi$ is simply considered as a constant free parameter, and $\xi=10^{-4}$ is taken for illustration. We should notice that, $\xi$ has very profound physical implications as it characterizes how the Universe is reheated. In further study, the value of $\xi$ should be computed in detail with a specific reheating process \cite{Bassett:2005xm, Allahverdi:2010xz, Amin:2014eta}. And then the strength of suppression of matter perturbation predicted by such reheating process can be obtained accordingly. In turn, future astrophysical observations on the strength of suppression can be also applied to constrain the cosmic reheating process.

\section{Acknowledgments} 
I thank Yeuk-kwan Edna Cheung, Xiaheng Xie, and Qing Chen for useful discussion, careful reading of this manuscript and important suggestion on improving the clarity of presentation. This work has been supported in parts by the National Natural Science Foundation of China (11603018, 11963005, 11775110, 11433004, 11690030), the Yunnan Provincial Foundation (No.2016FD006), and the Leading/Top Talents of Yunnan Province (2015HA022, 2015HA030).


\end{document}